\documentclass{svjour3}                     
\smartqed  
\usepackage{graphicx}
\journalname{General Relativity and Gravitation}

\newcommand{\as}{A_{\cal S}}
\newcommand{\bs}{B_{\cal S}}
\newcommand{\cs}{C_{\cal S}}
\newcommand{\ds}{D_{\cal S}}
\newcommand{\ax}{A_{\xi}}
\newcommand{\bx}{B_{\xi}}
\newcommand{\cx}{C_{\xi}}
\newcommand{\dx}{D_{\xi}}
\newcommand{\LN}{{\bf {\tilde L}}_{\rm N}}
\newcommand{\sx}{(4{\cal S}+3\fourvec{\xi})}
\newcommand{\ls}{({\bf {\tilde L_N}}\cdot{\cal S})}
\newcommand{\lx}{({\bf {\tilde L_N}}\cdot \fourvec{\xi})}
\newcommand{\lsx}{{\bf {\tilde L_N}}\cdot(4{\cal S}+3\fourvec{\xi})}
\newcommand{\ncs}{({\bf n}\cdot{\cal S})}
\newcommand{\nts}{{\bf n}\times{\cal S}}
\newcommand{\vcs}{({\bf v}\cdot{\cal S})}
\newcommand{\vts}{{\bf v}\times{\cal S}}
\newcommand{\ncx}{({\bf n}\cdot \fourvec{\xi})}
\newcommand{\ntx}{{\bf n}\times \fourvec{\xi}}
\newcommand{\vcx}{({\bf v}\cdot \fourvec{\xi})}
\newcommand{\vtx}{{\bf v}\times \fourvec{\xi}}

\newcommand{\es}{\epsilon^{7/2}}

\newcommand{\fourvec}[1]{\mbox{\boldmath $\mathsf{#1}$}}
\newcommand{\be}{\begin{equation}}
\newcommand{\ee}{\end{equation}}
\newcommand{\bea}{\begin{eqnarray}}
\newcommand{\eea}{\end{eqnarray}}
\newcommand{\bes}{\begin{subequations}}
\newcommand{\ees}{\end{subequations}}

\begin{document}

\title{Application of energy and angular momentum balance to gravitational
radiation reaction for binary systems with spin-orbit coupling
}


\author{Jing Zeng         \and
        Clifford M. Will 
}


\institute{
              McDonnell Center for the Space Sciences and Department of Physics \\
	      Washington University, St. Louis MO 63130 USA\\
              \email{cmw@wuphys.wustl.edu}             \\
              \email{jzeng@hbar.wustl.edu}             \\
}

\date{Received: date / Accepted: date}

\maketitle

\begin{abstract}
We study gravitational radiation reaction in the equations of motion for
binary systems with spin-orbit coupling, at order $(v/c)^7$ beyond
Newtonian gravity, or $O(v/c)^2$ beyond the leading radiation reaction
effects for non-spinning bodies.  We use expressions for the energy and
angular momentum flux at infinity that include spin-orbit corrections,
together with an assumption of energy and angular momentum balance, to
derive equations of motion that are valid for general orbits and for a class
of coordinate gauges.  We show that the equations of motion are compatible
with those derived earlier by a direct calculation.
\end{abstract}

\section{Introduction and Summary}
\label{sec:intro}

The backreaction of the emission of gravitational radiation on the system
emitting the radiation is a problem both of formal interest within general
relativity and of practical interest for gravitational-wave detection.  A
leading candidate source for laser-interferometric gravitational-wave
observatories, both on the ground and in space, is the radiation-reaction
induced inspiral of a binary
system of two compact objects (black holes or neutron stars).  In
order to develop accurate theoretical predictions for the gravitational
waveforms emitted by such systems, one must know
their evolution under the dissipative
effects of gravitational-wave emission to high accuracy.  In
addition, particularly for systems containing black holes, the effects of
spin may be important.  Spin-orbit and spin-spin couplings can result in
precessions of the bodies' spins and of the orbital angular momentum,
leading to modulations in the gravitational waveform \cite{3min}, 
and can affect the
rate of decay of the orbit \cite{kww,kidder}.  

As a result, substantial effort has gone into determining the effects of
spin in binary systems.  Except for the final few orbits, much of the
inspiral of such systems can be described by the  post-Newtonian
approximation, which is an expansion of Einstein's equations in powers of 
$\epsilon \sim (v/c)^2 \sim Gm/rc^2$, where $v$, $m$ and $r$ represent
typical velocities, masses and separations in the system, and $G$ and $c$
are the gravitational constant and speed of light.   Each power of
$\epsilon$ represents one
``post-Newtonian'' (PN) order in the series ($\epsilon^{1/2}$ represents
one-half, or 0.5PN orders). 

Formally, spin
effects first enter the equations of motion at the 1PN level, and
have been derived by
numerous authors from a variety of points of view, ranging from formal
developments of the GR equations of motion in multipole expansions
\cite{papapetrou1,papapetrou2}, to post-Newtonian calculations
\cite{obrien}, to
treatments of linearized GR as a spin-two quantum theory 
\cite{barkerocon1,barkerocon2}.
For a review of these various approaches, see \cite{barkeroconrev}.

Spin also affects gravitational radiation reaction, 
and radiation reaction can affect spin; it is straightforward to show that
such effects first occur at 3.5PN order.  In earlier work, we
derived, from first principles, the radiation-reaction 
effects of spin-orbit
and spin-spin coupling, by integrating the post-Newtonian 
hydrodynamic equations of motion,
including 1PN, 2.5PN and 3.5PN terms, over bodies consisting of rotating
fluid \cite{dire3,dire4}.  
As a check, we found that the loss of energy and angular momentum
(including spin) induced by the radiation reaction terms, matched precisely
the expressions for energy  and angular momemtum 
flux derived by Kidder {\em et
al.}\cite{kww,kidder}.  

An alternative approach to obtaining equations of motion with radiation
reaction at higher PN orders was studied by Iyer and Will
\cite{iyerwill1,iyerwill2}.   There, we wrote down the most general
form that the 2.5PN and 3.5PN radiation-reaction terms 
could take in the
equations of motion for a binary system of {\em spinless} bodies, in
terms of arbitrary coefficients. 
We then used the assumption of energy
and angular momentum balance, combined with 
energy and angular-momentum 
flux expressions accurate to PN order beyond the quadrupole approximation
to impose constraints on the arbitrary
coefficients used in the equations of motion.  
After taking into account a fundamental ambiguity in the definitions
of energy and angular momentum at 2.5PN and 3.5PN orders, we were left
with equations of motion with coefficients that are fixed up to two
arbitrary coefficients at 2.5PN order and 6 arbitrary coefficients at
3.5PN order.  It was then straightforward to show that these eight degrees
of freedom correspond precisely to the effects, mapped onto the two-body
equations of motion, of coordinate transformations at the relevant PN
orders.  At 2.5PN order, for example, one choice of the two arbitrary
coefficients gives the equations in the so-called Burke-Thorne gauge
\cite{MTW},
in which the radiation reaction terms are obtained from a gradient of
the potential $(G/5c^2) x^i x^j d^5 I^{<ij>}/dt^5$, where $I^{<ij>}$ is the
trace-free moment of inertia tensor of the system,
while another choice gives the Damour-Deruelle gauge, which is more directly
tied to harmonic gauge \cite{DD81,damour300}.
For spinless systems, this approach was extended to determine the 4.5PN
terms in the equations of motion using flux expressions accurate to 2PN
order beyond quadrupole \cite{gopuiyer2}.

It is the purpose of this paper to extend this approach to include
spin-orbit radiation reaction effects at 3.5PN order.  We assume that
the equation of motion for the relative vector 
${\bf x}={\bf x}_1-{\bf x}_2$ 
in a binary system may be written in a PN expansion in the form
\begin{equation}
{\bf a}= -\frac{m}{r^2}\bf n + {\bf a}_{\rm PN-SO}+ \dots +{\bf a}_{\rm 
2.5PN}+ \dots + {\bf a}_{\rm 3.5PN-SO} \,,
\label{PNeom}
\end{equation}
where $r \equiv |{\bf x}|$, 
${\bf n}\equiv {\bf x}/r$ and  
$m\equiv m_1+m_2$;
${\bf a}_{\rm PN-SO}$ is the post-Newtonian spin-orbit contribution, 
${\bf a}_{\rm 2.5PN}$ is the leading radiation-reaction 
contribution, and ${\bf a}_{\rm 3.5PN-SO}$ is the 3.5PN spin-orbit 
contribution. 
Here and for the rest of this paper, we use units in which $G=c=1$.
We have not displayed the point mass 1PN, 2PN, 3PN, and 3.5PN 
terms, as they will
play no role in our analysis.
We also will not use the bookeeping parameter $\epsilon$ explicitly to keep
track of PN orders, since we will be considering only specific orders.
It is sufficient to recall that, since spin scales as $mvr$, then $S/r^2
\sim v(m/r) \sim \epsilon^{3/2}$.  This, plus explicit labelling of terms
throughout, should make clear the PN order of terms being discussed.

We then write down the
most general 3.5PN spin-orbit expression
that (a) contains terms each involving a single spin (either ${\bf S}_1$ or
${\bf S}_2$), (b) is a vector, and (c) is antisymmetric under the
interchange $1 \rightleftharpoons 2$.  This turns out to involve 30
arbitrary coefficients.  

Because the bodies have intrinsic spin, we must make an assumption
about their spin evolution.  At 1PN order, we assume that they obey the
standard spin-orbit precession equations (see Eq. (\ref{eomPNSO}) below).  
These 1PN
equations produce only precession; the magnitudes of the spins do not
change.  At 3.5PN order, we likewise 
assume that gravitational radiation reaction 
produces only precessions of the spin.  This is a reasonable
assumption, because, for a rotating axisymmetric body, it is impossible
to see how gravitational radiation can cause it to spin up or down, to the
3.5PN order being considered.
Such effects must involve specific couplings of radiation to deformations of
the bodies, either due to rotational flattening or due to tidal couplings, 
which
are beyond the scope of our assumption of almost point-like bodies with
spin.
In this case, we can then show that the most general 3.5PN expression for
the evolution of each spin that (a) is a pseudovector; (b) depends
only on the spin itself and on orbital variables; and 
(c) is orthogonal to the spin, can in fact be written as a total time
derivative of spin and orbital variables, which can then be absorbed into a
meaningless 3.5PN-order correction to the definition of the spin.  

Consequently, we can calculate the loss of energy and angular 
momentum using only the parametrized equations of motion and the 1PN spin
precession equations.  There is no contribution to the evolution of the
spins at 3.5PN order.  However, we must incorporate the freedom to add
arbitrary terms of 3.5PN spin-orbit order into the definitions of 
total energy and total
angular momentum, just as in the spinless case.  
There are 6 such terms in $E$ and 26 in ${\bf J}$.  Thus
there is a total of 
62 arbitrary coefficients to be determined.  We then equate the
time derivative of these expressions for $E$  and ${\bf J}$ with
the corresponding expressions obtained from  
the far-zone gravitational-wave flux, including spin-orbit
terms, as calculated by Kidder {\em et al.}\cite{kww,kidder}, and compare
them term by term.  This leads to 54 constraints on the coefficients; however
4 of these constraints are not linearly independent of others, and thus we
have 50 constraints on 62 coefficients, leaving 12 
undetermined coefficients.  Finally we show that these 12
free coefficients in the equation of motion correspond precisely to the
effects of 3.5PN order coordinate transformations, mapped onto the two-body
equations
of motion with spin-orbit coupling.

The remainder of this paper provides details.
In Sec. \ref{sec:two body} we review the known equations of motion and spin
evolution through 2.5PN order.  Section \ref{sec:balance} applies energy and
angular momentum balance to determine the 3.5PN spin-orbit terms in the
two-body equations of motion, while Sec. \ref{sec:gauge} shows that the
remaining undetermined coefficients are directly related to gauge freedom.
Section \ref{sec:conclusions} presents concluding remarks, while certain
detailed formulae are relegated to Appendices.

\section{Two-body equations of motion with spin-orbit 
coupling}
\label{sec:two body}

The PN-SO and 2.5PN terms in Eq. (\ref{PNeom}) are given by
conventional expressions
\begin{eqnarray}
{\bf a}_{\rm PN-SO}  &=& \frac{1}{r^3} \biggl \{
\frac{3}{2} \frac{\bf n}{r} {\bf {\tilde L}}_{\rm N} \cdot
\left ( 4{\bf {\cal S}} + 3\fourvec{\xi} \right )
- {\bf v} \times
\left ( 4{\bf {\cal S}} + 3\fourvec{\xi} \right )
+\frac{3}{2} \dot r {\bf n} \times
\left ( 4{\bf {\cal S}} + 3\fourvec{\xi} \right )
\biggr \} \,,
\label{eomPNSO}
\\
{\bf a}_{\rm 2.5PN}&=&\frac{8\mu m}{5r^3}\biggl 
\{\left [3(1+\beta)v^2+\frac{1}{3}(23+6\alpha-
9\beta)\frac{m}{r}
-5\beta \dot r^2\right ]\dot r {\bf n}
\nonumber
\\
&&-\left [(2+\alpha)v^2+(2-\alpha)\frac{m}{r}-3(1+\alpha)\dot 
r^2\right ]
{\bf v}\biggr \} \,,
\label{eom25PN}
\end{eqnarray}
where ${\bf v}=d{\bf x}/dt$ is the relative velocity, $\mu 
\equiv m_1m_2/m$ is the reduced mass,
${\cal S}\equiv {\cal S}_1+{\cal S}_2$ is the total spin, 
$\fourvec{\xi}=(m_2/m_1){\cal S}_1+(m_1/m_2){\cal S}_2$ is a second spin 
parameter, 
$\LN ={\bf x}\times {\bf v}$ is the orbital angular momentum 
per unit reduced
mass, and $\dot r = {\bf v}\cdot {\bf n}$.

The coefficients $\alpha$ and $\beta$ in ${\bf a}_{\rm 2.5PN}$ reflect the
possibility of different gauges for expressing radiation reaction at 2.5PN
order
\cite{iyerwill1,iyerwill2}.  The
choice $\alpha=4$, $\beta=5$ corresponds to Burke-Thorne
gauge \cite{MTW}, 
while the choice $\alpha=-1$,
$\beta=0$ leads to the Damour-Deruelle radiation-reaction formula
\cite{DD81,damour300}.  Any choice of $\alpha$ and $\beta$ leads to the same
loss of energy and angular momentum at 2.5PN order, corresponding to
quadrupole approximation energy and angular momentum flux.

In defining spin, we must specify the center of mass of each 
body using a
procedure commonly known as the ``spin supplementary 
condition (SSC); the definition used in this paper
corresponds to the value $k_{\rm SSC} =1/2$ (see
\cite{barkeroconssc,kidder,dire3} for further discussion).  

The equations of evolution for the spins may be written 
in the form
\be
{\dot {\bf {\cal S}}}_1 = ({\dot {\bf {\cal S}}}_1)_{\rm PN} +\dots + 
({\dot {\bf {\cal S}}}_1)_{\rm 3.5PN-SO} \,,
\ee
where
\be
({\dot {\bf {\cal S}}}_1)_{\rm PN} = 
\frac{\mu}{r^3} {\bf {\tilde L}}_{\rm N}
\times {\bf {\cal S}}_1 \left ( 2 + \frac{3}{2}\frac{m_2}{m_1} \right ) \,,
\label{spinsummary}
\ee
and where the equations for spin 2 can be obtained by the interchange $1
\rightleftharpoons 2$.  We have not included conservative 2PN and 3PN
contributions, and can show that the leading radiation reaction contributions
come at 3.5PN order \cite{dire3}.

Up to 2.5PN order, the motion is conservative and the energy 
and angular momentum are constant.  Including only the Newtonian and spin-orbit
terms, they are given by
\begin{eqnarray}
E&=& E_{\rm N} = \mu \biggl(\frac{1}{2}v^2-\frac{m}{r}\biggr) \,,
\label{EJ1}
\\
{\bf J}&=&\mu {\tilde {\bf L}}_{\rm N}+ {\bf {\cal S}}+\frac{\mu}{2r} {\bf n} \times 
\biggl[ {\bf n} \times \sx 
\biggr] \,.
\label{EJ2}
\end{eqnarray}
In our chosen spin 
supplementary condition, there is no spin-orbit contribution 
to the conserved energy, while ${\bf J}$ contains the orbital angular
momentum, the total spin, and a PN spin-orbit contribution.
These conserved quantities can be derived from the 
equations of motion by constructing 
$\frac{1}{2}dv^2/dt\equiv {\bf v}\cdot {\bf a}$, 
and $d({\bf x}\times {\bf v})/dt\equiv {\bf x}\times {\bf a}$, 
and showing 
that, after substituting the equations of motion and spin-precession 
equations carried 
to the appropriate order, everything can be 
expressed as total time derivatives.

\section{Spin-orbit radiation reaction via $E$ and $J$ balance}
\label{sec:balance}

We now write down the most general 3.5PN spin-orbit terms as
\begin{eqnarray}
    {\bf a}_{3.5PN-SO}&=&-\frac{\mu}{5r^4}\biggl[\as \frac{\dot 
r{\bf n}}{r}\ls+\bs \frac{\bf v}{r}\ls+\cs \dot r\vts +\ds \nts
    \nonumber
    \\
    &&+\ax \frac{\dot r{\bf n}}{r}\lx+\bx \frac{\bf v}{r}\lx
    +\cx \dot r\vtx +\dx \ntx\biggr] \,.
\label{eomgeneral}
\end{eqnarray}
The form of Eq. (\ref{eomgeneral})
is dictated by the fact that it must be a correction 
to the Newtonian acceleration, ({\em i.e.} be proportional to 
a mass $/r^2$); must vanish in the test body limit when 
gravitational radiation vanishes, ({\em i.e.} be proportional to 
$\mu$); must be dissipative, or odd in velocities; 
must be linear in the  
spins; must be a vector, not a pseudovector; and must change sign under
the interchange $1 \rightleftharpoons 2$.  Note that other possible terms,
such as $\LN \ncs$ can be seen to be linear combinations of the terms above
using standard vector identities.  The 
prefactor 1/5 is chosen for convenience. To make the 
terms of $O(\es)$ beyond Newtonian order, $\as$, $\bs$, $\cs$, 
$\ax$, $\bx$ and $\cx$ must be of $O(\epsilon)$, and $\ds$ and $\dx$ 
must be of $O(\epsilon^2)$.  
The only orbital variables available to construct expressions
of the relevant order are 
$v^2$, $m/r$ and $\dot r^2$. Thus $\as$, $\bs$, 
$\cs$ and $\ds$ can be written in terms of 15 arbitrary coefficients, in the
form
\begin{eqnarray}
\as&=&a_1v^2+a_2\frac{m}{r}+a_3\dot r^2 \,,
\nonumber \\
\bs&=&a_4v^2+a_5\frac{m}{r}+a_6\dot r^2 \,,
\nonumber \\
\cs&=&a_7v^2+a_8\frac{m}{r}+a_9\dot r^2 \,,
\nonumber \\
\ds&=&a_{10}v^4+a_{11}v^2\dot r^2+a_{12}\dot 
r^4+a_{13}v^2\frac{m}{r}+a_{14}\dot 
r^2\frac{m}{r}+a_{15}\frac{m^2}{r^2} \,.
\label{coefficients}
\end{eqnarray}
In a parallel manner, we can write $\ax$, $\bx$,
$\cx$ and $\dx$ in terms of its own set of 15 coefficients.
Because all expressions involving spin-orbit terms divide naturally
into those involving the total spin ${\bf {\cal S}}$ and those
involving the spin parameter $\fourvec{\xi}$, we can solve for each
set using identical methods; we will focus on the ${\bf {\cal
S}}$-terms.
Our goal is to evaluate these thirty coefficients by imposing energy and
angular momentum balance.  

Because the equations of motion at 2.5PN order and 3.5PN 
order have dissipative terms, the energy and angular momentum 
are no longer conserved explicitly.  Furthermore, they are 
ambiguous because one has the freedom to add arbitrary terms 
to $E$ and $\bf J$ at 2.5PN order and 3.5PN order to redefine them 
without affecting their conservation through 2PN order.  
Similarly, the spins are strictly
defined only up to the order at which radiation
reaction begins, and so one has the freedom to add a 3.5PN term to each
spin, without changing its behavior at 	``conservative'' orders.

Adding to $E$ and $\bf J$ the appropriate 2.5PN terms to account for the
coefficients $\alpha$ and $\beta$ in Eq. (\ref{eom25PN}), and adding 
the most general 3.5PN spin-orbit terms, with arbitrary
coefficients, we can define new quantities $E^*$, and $\bf J^*$
according to
\bea
E^* &=& E_{\rm N} + \delta E_{\rm 2.5PN} + \delta E_{\rm 3.5PN-SO} \,,
\nonumber \\
\bf J^* &=& \mu {\tilde {\bf L}}_{\rm N}   
 + {\bf {\cal S}}+\frac{\mu}{2r} {\bf n} \times
 \left[ {\bf n} \times \sx \right] 
 +\delta  {\bf J}_{\rm 2.5PN} + \delta {\bf J}_{\rm 3.5PN-SO} \,,
\label{EJstar}
\eea
where, from our earlier work \cite{iyerwill1,iyerwill2}, we can write
\bea
\delta E_{\rm 2.5PN}&=&  \frac{8}{5} 
\frac{\mu^2 m}{r^2} {\dot r} [(2+\alpha)v^2 - \beta {\dot r}^2 ] \,, 
\nonumber \\
\delta {\bf J}_{\rm 2.5PN} &=& \alpha \frac{8}{5} 
\frac{\mu^2 m}{r^2}{\dot r} \LN
\,,
\eea
and for the 3.5PN-SO expressions we write the general parametrized form
\bea
\delta E_{\rm 3.5PN-SO}&=&  - \frac{1}{5} \frac{\mu^2}{r^4}{\dot r} 
\left \{ \ls \left ( \alpha_1 v^2
+ \alpha_2 {\dot r}^2 + \alpha_3 \frac{m}{r} \right ) 
+ ({\cal S} \to \fourvec{\xi} ) \right \} \,,
\nonumber \\
\delta {\bf J}_{\rm 3.5PN-SO} &=& -\frac{1}{5} \frac{\mu^2}{r^2} \biggl \{
{\dot r} {\bf {\cal S}} \left ( \gamma_1 v^2 + \gamma_2 {\dot r}^2 +
\gamma_3 \frac{m}{r} \right )
+ {\dot r} {\bf n} ({\bf n} \cdot {\bf {\cal S}})
\left ( \gamma_4 v^2 + \gamma_5 {\dot r}^2 +
\gamma_6 \frac{m}{r} \right )
\nonumber \\
&& + {\bf v} ({\bf n} \cdot {\bf {\cal S}})
\left ( \gamma_7 v^2 + \gamma_8 {\dot r}^2 +
\gamma_9 \frac{m}{r} \right )
+ {\bf n}({\bf v} \cdot {\bf {\cal S}})
\left ( \gamma_{10} v^2 + \gamma_{11} {\dot r}^2 +
\gamma_{12} \frac{m}{r} \right )
\nonumber \\
&&+ \gamma_{13} {\dot r} {\bf v}({\bf v} \cdot {\bf {\cal S}})
+ ({\cal S} \to \fourvec{\xi} ) 
\biggr \}\,,
\label{EJstar35}
\eea
where the notation ${\cal S} \to \fourvec{\xi}$ means repeat the preceding
terms replacing ${\cal S}$ with $\fourvec{\xi}$, with an appropriate
set of arbitrary
coefficients.  This
gives a total of 32 arbitrary coefficients.

We now take time 
derivatives of $E^*$ and ${\bf J}^*$ in Eqs. (\ref{EJstar}), 
substituting the Newtonian and PN spin-orbit accelerations explicitly, 
to obtain
\begin{eqnarray}
{\dot E}^*&=&\mu {\bf v}\cdot \left ({\bf a}_{\rm 2.5PN}+
{\bf a}_{\rm 3.5PN-SO} \right ) + \frac{d}{dt} \delta E_{\rm 2.5PN}
+ \frac{d}{dt} \delta E_{\rm 3.5PN-SO} \,,
\nonumber
\\
{\dot {\bf J}}^*&=& {\dot {\bf {\cal S}}}_{\rm 3.5PN-SO}
+ \mu {\bf x}\times \left ( {\bf a}_{\rm 2.5PN}+
{\bf a}_{\rm 3.5PN-SO} \right ) + \frac{d}{dt} \delta {\bf J}_{\rm 2.5PN}
+ \frac{d}{dt} \delta {\bf J}_{\rm 3.5PN-SO} \,.
\end{eqnarray}

In fact, we will show in Appendix \ref{app:spinevolve} that, if we 
assume that $({\dot {\bf {\cal S}}}_1 )_{\rm 3.5PN-SO}$ is orthogonal to
${\bf {\cal S}}_1$ (and similarly for spin 2), 
then the most general 3.5PN expression for $({\dot {\bf {\cal S}}}_1 )_{\rm
3.5PN-SO}$ turns out to be a total time derivative, which can be absorbed into
a meaningless 3.5PN correction to the definition of ${\bf {\cal S}}_1$.
Hence we can assume 
henceforth that ${\dot {\bf {\cal S}}}_{\rm 3.5PN-SO} =({\dot {\bf
{\cal S}}}_1 )_{\rm 3.5PN-SO} + ({\dot {\bf {\cal S}}}_2 )_{\rm
3.5PN-SO} =0$.

We now substitute the appropriate terms from the equations of motion
(\ref{eom25PN}) and (\ref{eomgeneral}), 
and calculate explicitly the time derivatives of the 2.5PN and
3.5PN-SO contributions to $E^*$ and ${\bf J}^*$.  These time-derivative
terms may be
calculated using the identities shown in Appendix \ref{app:timederiv},
which are derived using the Newtonian equations of motion and the 1PN
spin-orbit terms.  When evaluating $d\delta {E}_{\rm 2.5PN}/dt$ and
$d\delta {\bf J}_{\rm 2.5PN}/dt$, in order 
to obtain all terms that contribute at 3.5PN-SO order, we
must include the 1PN spin orbit terms present in the expressions in Appendix
\ref{app:timederiv}.  The result is 
\begin{eqnarray}
\dot E^*&=&-\frac{8\mu^2m^2}{15r^4}(12v^2-11\dot r^2)-
\frac{8\mu^2m}{10r^6}\lsx \biggl[-(2+\alpha)v^2+3\beta \dot
r^2\biggr]
\nonumber
\\
&&- \frac{\mu^2}{5r^5} \biggl \{ \ls \left (
{\cal P}_1 v^4 + {\cal P}_2 v^2 {\dot
r}^2 + {\cal P}_3 {\dot r}^4 + {\cal P}_4 v^2 \frac{m}{r}
+ {\cal P}_5 {\dot r}^2 \frac{m}{r} + {\cal P}_6 \frac{m^2}{r^2} 
\right ) 
\nonumber
\\
&& \quad + ({\bf {\cal S}} \to \fourvec{\xi} ) \biggr \}\,,
\label{Estardot}
\\
\dot J^*&=&-\frac{8\mu^2m}{5r^3} \LN \left (2v^2-3\dot
r^2+2\frac{m}{r} \right )
\nonumber
\\
&&-\frac{8\mu^2m\alpha}{5r^4}\biggl \{-\frac{1}{2r^2}\LN
\lsx+\dot r{\bf n}\times \biggl[({\bf v}-\frac{3}{2}\dot r{\bf n})\times \sx
\biggr]\biggr \}
\nonumber
\\
&& -\frac{\mu^2}{5r^3} \biggl \{ 
{\bf {\cal S}}
\left ({\cal R}_1 v^4 + {\cal R}_2 v^2 {\dot
r}^2 + {\cal R}_3 {\dot r}^4 + {\cal R}_4 v^2 \frac{m}{r}
+ {\cal R}_5 {\dot r}^2 \frac{m}{r} + {\cal R}_6 \frac{m^2}{r^2}
\right ) 
\nonumber
\\
&&
+{\bf n}({\bf n}\cdot {\bf {\cal S}})
 \left ({\cal R}_7 v^4 + {\cal R}_8 v^2 {\dot
r}^2 + {\cal R}_9 {\dot r}^4 + {\cal R}_{10} v^2 \frac{m}{r}
+ {\cal R}_{11} {\dot r}^2 \frac{m}{r} + {\cal R}_{12} \frac{m^2}{r^2} 
\right )
\nonumber
\\
&&
+{\dot r} {\bf n}({\bf v}\cdot {\bf {\cal S}}) 
\left ( {\cal R}_{13} v^2 + {\cal R}_{14} {\dot r}^2  
+ {\cal R}_{15} \frac{m}{r} \right )
+ {\dot r} {\bf v}({\bf n}\cdot {\bf {\cal S}})
 \left ( {\cal R}_{16} v^2 + {\cal R}_{17} {\dot r}^2  + 
{\cal R}_{18} \frac{m}{r} \right)
\nonumber
\\
&&
+ {\bf v}({\bf v}\cdot {\bf {\cal S}})
\left ( {\cal R}_{19} v^2 + {\cal R}_{20} {\dot r}^2  + {\cal R}_{21}
\frac{m}{r} \right )
 + ({\cal S} \to \fourvec{\xi} ) \biggr \}\,.
\label{Jstardot}
\end{eqnarray}
The first term in each of Eqs. (\ref{Estardot}) and (\ref{Jstardot}) is 
the 2.5PN quadrupole, or
Newtonian loss term, while the second term in each case
comes from the spin-orbit
correction terms in Appendix \ref{app:timederiv} applied to 
$d \delta E_{\rm 2.5PN}/dt$ and $d \delta {\bf J}_{\rm 2.5PN}/dt$.  In
the third set of terms in each case,
the 27 coefficients ${\cal P}_n, \, n=1 \dots 6$ and 
${\cal R}_n, \, n=1 \dots 21$ in the ${\cal S}$-dependent terms
are functions of the 15 
coefficients $a_n$ from the equations of motion (\ref{eomgeneral}) and 
(\ref{coefficients}), and of the 
16 coefficients $\alpha_n$ and $\gamma_n$ from the 3.5PN
ambiguity terms in $E^*$
and ${\bf J}^*$.  A parallel set of 27 coefficients appear in the
$\fourvec{\xi}$-dependent terms, with identical dependences on the
corresponding 15 + 16 arbitrary coefficients.

We now use the assumption of energy and angular momentum 
balance to equate the rate of energy and angular momentum 
loss to the corresponding far-zone fluxes \cite{kww,kidder}. 
The lowest-order Newtonian and the 1PN spin-orbit contributions are 
given by 
\begin{eqnarray}
{\dot E}_{\rm far\, zone}&=&-
\frac{8}{15}\frac{\mu^2m^2}{r^4}(12v^2-11\dot r^2)
\nonumber 
\\
&&- \frac{8\mu^2m}{15r^6}\biggl \{ \ls \left ( 27\dot r^2-37v^2-
12\frac{m}{r} \right )
\nonumber
\\
&&
+\lx \left ( 18\dot r^2-19v^2-8\frac{m}{r} \right )\biggr \} \,,
\label{farfluxE}
\\
{\dot {\bf J}}_{\rm far\, zone}&=&-
\frac{8}{5}\frac{\mu^2m}{r^3}\LN (2v^2-3\dot r^2+2\frac{m}{r})
\nonumber 
\\
&&-\frac{4\mu^2}{5r^3}\biggl \{ {\bf {\cal S}} \left (
6v^2\dot r^2
- 6v^4 
- \frac{50}{3}v^2\frac{m}{r}
+\frac{50}{3}\dot r^2 \frac{m}{r}
-2\frac{m^2}{r^2}
\right )
\nonumber
\\
&&+{\bf n} \ncs 
\left (
18v^4
-30\dot r^2 v^2
+25v^2 \frac{m}{r}
+6\dot r^2 \frac{m}{r}
+2\frac{m^2}{r^2}
\right )
\nonumber
\\
&&+\dot r {\bf n} \vcs \left ( 6v^2-21\frac{m}{r} \right )
- \dot r {\bf v} \ncs \left (18v^2-30\dot r^2 +33\frac{m}{r} \right )
\nonumber
\\
&&+ {\bf v}\vcs \left ( 6v^2-12\dot r^2+23\frac{m}{r} \right )
\nonumber
\\
&&+ \fourvec{\xi} \left (
5\dot r^4
-2v^2 \dot r^2 
- \frac{10}{3}v^4
-\frac{22}{3}v^2 \frac{m}{r}
+\frac{23}{3}\dot r^2 \frac{m}{r}
- \frac{4m^2}{3r^2} 
\right )
\nonumber
\\
&&+ {\bf n} \ncx \left (
13v^4
-20\dot r^2 v^2
+\frac{41}{3}v^2 \frac{m}{r}
+6\dot r^2 \frac{m}{r}
+\frac{4m^2}{3r^2}
\right )
\nonumber
\\
&&+\dot r {\bf n} \vcx \left (7v^2-5\dot r^2-\frac{34m}{3r} \right )
-\dot r {\bf v} \ncx \left (13v^2-20\dot r^2 +\frac{64m}{3r} \right )
\nonumber
\\
&&+ {\bf v} \vcx \left (\frac{10}{3}v^2- 5\dot r^2 +\frac{38m}{3r}\right ) 
\biggr \} \,.
\label{farfluxJ}
\end{eqnarray}

After rewriting some of the terms in Eq. (\ref{Jstardot}) using standard
vector identities, we compare Eqs. (\ref{Estardot}) and
(\ref{Jstardot}) to Eqs. (\ref{farfluxE}) and (\ref{farfluxJ}) term
by term to obtain 54 constraints on the 62 coefficients.  It turns out,
however, that 4 of these constraints are not linearly independent of
others, so there are 50 non-trivial constraints, leaving 12
undetermined degrees of freedom.  The specific choice of the free
coefficients is somewhat arbitrary; one choice gives the following values
for the coefficients (\ref{coefficients})
in the equations of motion (\ref{eomgeneral}):
\begin{eqnarray}
a_1&=&2820[2160]+15\gamma_4+45\gamma_7+45\gamma_9
+15\gamma_{11}+45\gamma_{12}-3\alpha_2
\,,
\nonumber
\\
a_2&=&-1728[-1348]-13\gamma_4-39\gamma_7-42\gamma_9-11\gamma_{11}-
42\gamma_{12}+3\alpha_2
\,,
\nonumber
\\
&& +48[36](\alpha-\beta)
\,,
\nonumber
\\
a_3&=&-6020[-4620]-35\gamma_4-105\gamma_7-105\gamma_9-35\gamma_{11}-
105\gamma_{12}+7\alpha_2
\,,
\nonumber
\\
a_4&=&-220[-164]-\gamma_4-3\gamma_7-3\gamma_9-2\gamma_{11}-3\gamma_{12}
\,,
\nonumber
\\
a_5&=&\frac{68}{3}[36]+\gamma_4+3\gamma_7+3\gamma_9+2\gamma_{11}
+3\gamma_{12}+16[12]\alpha
\,,
\nonumber
\\
a_6&=&860[640]+5\gamma_4+15\gamma_7+15\gamma_9+10\gamma_{11}+15\gamma_{12}
\,,
\nonumber
\\
a_{7}&=&-788[-608]-4\gamma_4-6\gamma_7-15\gamma_9-4\gamma_{11}-
15\gamma_{12}
\,,
\nonumber
\\
a_{8}&=&\frac{3152}{3}[808]+4\gamma_4+16\gamma_7+24\gamma_9+4\gamma_{11}+16\gamma_{12}-32[-24]\alpha
\,,
\nonumber
\\
a_{9}&=&2460[1900]+10\gamma_4+30\gamma_7+45\gamma_9+10\gamma_{11}+45\gamma_{12}
\,,
\label{solution}
\nonumber
\\
a_{10}&=&-148[-112]-2\gamma_4-3\gamma_7-3\gamma_9-2\gamma_{11}-3\gamma_{12}
\,,
\nonumber
\\
a_{11}&=&3320[2540]+25\gamma_4+60\gamma_7+60\gamma_9+25\gamma_{11}+60\gamma_{12}
\,,
\nonumber
\\
a_{12}&=&-6020[-4620]-35\gamma_4-105\gamma_7-105\gamma_9-35\gamma_{11}-
105\gamma_{12}
\,,
\nonumber
\\
a_{13}&=&\frac{1276[968]}{3}+4\gamma_4+11\gamma_7+9\gamma_9+4\gamma_{11}+5\gamma_{12}
\,,
\nonumber
\\
a_{14}&=&-4392[-3372]-23\gamma_4-87\gamma_7-78\gamma_9-23\gamma_{11}-
54\gamma_{12}+48[36]\alpha
\,,
\nonumber
\\
a_{15}&=&-376 \left [-\frac{872}{3} \right ]-2\gamma_4-8\gamma_7-6\gamma_9-2\gamma_{11}-2\gamma_{12}
\,,
\end{eqnarray}
where the numbers in square brackets represent the values to be used,
along with the corresponding set of six free coefficients, for the terms
in Eq. (\ref{eomgeneral}) involving $\fourvec{\xi}$.  

The unique choice of the twelve coefficients
\be
\alpha_2=\frac{45}{2}
,\,
\gamma_4=\frac{287}{2}
,\,
\gamma_7=-\frac{89}{6}
,\,
\gamma_9=-\frac{140}{3}
,\,
\gamma_{11}=-\frac{263}{2}
,\,
\gamma_{12}=-1
,\,
\ee
for the ${\bf {\cal S}}$ terms, and
\be
\alpha_2=-\frac{105}{2}
,\,
\gamma_4=\frac{181}{2}
,\,
\gamma_7=-\frac{155}{6}
,\,
\gamma_9=-34
,\,
\gamma_{11}=-\frac{105}{2}
,\,
\gamma_{12}=2 \,,
\ee
for the $\fourvec{\xi}$ terms, along with the values $\alpha=-1$,
and $\beta=0$ for 
the harmonic Damour-Deruelle gauge at 2.5PN order, gives precisely the
3.5PN spin-orbit radiation reaction terms derived in \cite{dire3}. 

\section{Gauge Freedom and Arbitrary Coefficients in the Equation of Motion}
\label{sec:gauge}

The formulas for energy and angular momentum flux in the far
zone are gauge invariant,
while the equations of motion are 
gauge, or coordinate dependent.  Any coordinate transformation $x^\mu
\to x^\mu + \zeta^\mu$, where $\zeta^\mu$ is, in a suitable sense, of 2.5PN
and 3.5PN order relative to $x^\mu$, will induce changes in the
variables of a binary system, such as the relative vector ${\bf x}$
and the spin vectors.  Notice that a transformation of coordinate time
simply induces a velocity-dependent change in ${\bf x}$ via 
${\bf x}(t+\delta t) = {\bf x}(t) + {\bf v}\delta t$.  
As for the spin, any change induced by a gauge transformation at 2.5PN
or 3.5PN order can always be reabsorbed into a new definition of spin,
since it is ambiguous at radiation-reaction orders.  
Therefore we will only consider coordinate transformation induced
changes in the relative vector ${\bf x}$ at 2.5PN and
3.5PN-SO orders, according to
\be
{\bf x}^\prime = {\bf x} + \delta {\bf x}_{\rm 2.5PN}
+\delta {\bf x}_{\rm 3.5PN-SO} \,.
\ee
The 2.5PN order coordinate change that corresponds to the arbitrary
coefficients $\alpha$ and $\beta$ in Eq. (\ref{eom25PN}) was 
calculated in \cite{iyerwill1,iyerwill2}, and is given by
\be
\delta {\bf x}_{\rm 2.5PN} = \frac{8\mu m}{15r}\left [\beta \dot r 
{\bf n}+(2\beta-3\alpha) {\bf v}\right ] \,.
\label{dx25PN}
\ee
We can derive directly
\begin{eqnarray}
{\bf v}'&=&{\bf v}+\delta {\dot {\bf x}}_{\rm 2.5PN}
+ \delta {\dot {\bf x}}_{\rm 3.5PN-SO} \,,
\nonumber
\\
\frac{d{\bf v}'}{dt'}&=&\frac{d{\bf v}}{dt}+\delta {\ddot {\bf
x}}_{\rm 2.5PN}
+ \delta {\ddot {\bf x}}_{\rm 3.5PN-SO} \,,
\nonumber
\\
\frac{m{\bf x}'}{r'^3}&=&\frac{m{\bf x}}{r^3}+\frac{m}{r^3}
(\delta {\bf x}_{\rm 2.5PN} -3{\bf n}{\bf n}\cdot \delta {\bf x}_{\rm
2.5PN}) 
\nonumber
\\ 
&&+ \frac{m}{r^3}
(\delta {\bf x}_{\rm 3.5PN-SO} -3{\bf n}{\bf n}\cdot \delta {\bf x}_{\rm
3.5PN-SO})  \,.
\end{eqnarray}
The 2.5PN terms in these equations must also be used to determine the
induced change in the 1PN spin-orbit acceleration terms in Eq.
(\ref{eomPNSO}). 
In evaluating $\delta {\ddot {\bf x}}_{\rm 2.5PN}$ explicitly using Eq.
(\ref{dx25PN}), the
1PN spin-orbit equations must be employed wherever an acceleration
occurs.
The result is that the equation of motion (\ref{PNeom}) changes
between the original and the new coordinates by a
quantity ${\bf Q}$ given by
\begin{eqnarray}
{\bf Q}&=&\biggl \{\frac{8\mu m}{5r^3}\biggl[\left (3\beta 
v^2+(2\alpha-3\beta)\frac{m}{r}-5\beta \dot r^2\right )\dot r {\bf n}
- \left (v^2-\frac{m}{r}-3\dot r^2 \right )\alpha {\bf v} \biggr]\biggr \}
\nonumber
\\
&& - \delta {\ddot {\bf x}}_{\rm 3.5PN-SO}  
- \frac{m}{r^3}
(\delta {\bf x}_{\rm 3.5PN-SO} -3{\bf n}{\bf n}\cdot \delta {\bf
x}_{\rm 3.5PN-SO})
\nonumber
\\
&&- \frac{8\mu m}{5r^5}\biggl[ \frac{1}{2r}\lsx
\left (3 (\alpha- \beta) {\dot r} {\bf n}
+ \alpha {\bf v} \right )
- \dot r {\bf v} \times \sx \left ( \alpha + \frac{1}{3} \beta \right )
\nonumber
\\
&&-\frac{1}{6} {\bf n} \times \sx \left (\beta v^2 -
\beta \frac{m}{r}-(9\alpha+6\beta)\dot r^2 \right) 
\biggr]
\,.
\label{Qterm}
\end{eqnarray}
Note that the 2.5PN terms in Eq. (\ref{Qterm}) match exactly the
arbitrary terms in Eq. (\ref{eom25PN}).
We now want to find a form for $\delta {\bf x}_{\rm 3.5PN-SO}$ so that
the 3.5PN-SO terms in Eq. (\ref{Qterm}) match the terms in
(\ref{eomgeneral})
generated by the arbitrary coefficients in Eq. (\ref{coefficients}).  
This can be
done either by direct integration to find $\delta {\bf x}_{\rm
3.5PN-SO}$, or by assuming a suitable form for $\delta {\bf x}_{\rm
3.5PN-SO}$ and seeing if one can solve for a set of coefficients.
Remarkably, a solution can be found, and is given by
\begin{eqnarray}
\delta {\bf x}_{\rm 3.5PN-SO}&=&
- \frac{\mu}{5r^2} \biggl \{
\frac{{\dot r}{\bf n}}{r} \ls \left (\gamma_4 + 3\gamma_7 + 3\gamma_9 +
\gamma_{11} + 3\gamma_{12} -\frac{1}{5}\alpha_2 \right ) 
\nonumber 
\\
&&
+ \frac{\bf v}{3r} \ls \left ( \gamma_4 + 3\gamma_7 + 3\gamma_9 +
3\gamma_{12} - \frac{1}{5}\alpha_2 \right )
\nonumber 
\\
&&
+ \frac{1}{4} \nts \biggl [ (\gamma_7+\gamma_9+\gamma_{12} )v^2
+ (\gamma_4 + 3\gamma_7 + 3\gamma_9 +
\gamma_{11} + 3\gamma_{12} ) {\dot r}^2 
\nonumber
\\
&&
- (\gamma_{12}-\frac{4}{3}\beta) \frac{m}{r} \biggr ]
- \frac{1}{4} {\dot r} \vts (\gamma_9 + \gamma_{12} ) 
 + ({\cal S} \to \fourvec{\xi} ) \biggr \}\,.
\end{eqnarray}
The 12 ($6+6$) coefficients correspond precisely to the 12 degrees of
freedom in Eqs. (\ref{solution}).

\section{Concluding remarks}
\label{sec:conclusions}

We have used energy and angular momentum balance to deduce the general form
of the 3.5PN spin-orbit radiation reaction terms in the two-body equations
of motion, and showed that the remaining undetermined degrees of freedom
correspond to the freedom to change gauges or coordinates at the
corresponding post-Newtonian order.  A specific choice of the free
coefficients yields 3.5PN spin-orbit terms in the 
equations of motion identical with those derived from
first principles.
The results were subject to the physically reasonable assumption that
gravitational radiation reaction has no effect on the magnitude of the
individual spins, to 3.5PN order.    

A natural extension of this work is to determine the contribution of spin-spin
interactions in radiation reaction using balance arguments and to compare
the results with those calculated from first principles by
Wang and Will \cite{dire4}.  This work is in progress.

\appendix
\section{Extracting total time derivatives}
\label{app:timederiv}

Using the Newtonian equations of motion plus the 1PN spin-orbit terms, it is
straightforward to establish a number of  identities, which may be used to
relate collections of terms to 
total time derivatives of other expressions. 
For any non-negative integers $s$, $p$ and $q$, we obtain
\bea
\frac{d}{dt} \left ( \frac{v^{2s} {\dot r}^p}{r^q} \right ) &=&
\frac{v^{2s-2} {\dot r}^{p-1}}{r^{q+1}}
\biggl \{ pv^4 - (p+q)v^2{\dot r}^2 - 2s{\dot r}^2\frac{m}{r}
-pv^2\frac{m}{r} 
\nonumber
\\
&&
+ \frac{p}{2}\frac{v^2}{r^3} {\bf {\tilde L}}_{\rm N} \cdot
(4 {\bf {\cal S}} + 3 \fourvec{\xi}) \biggr \}
\,,
\nonumber
\\
\frac{d}{dt} \left ( \frac{v^{2s} {\dot r}^p}{r^q} {\bf {\tilde L}}_{\rm N}
\right ) &=&
{\bf {\tilde L}}_{\rm N}\frac{d}{dt} \left ( \frac{v^{2s} {\dot
r}^p}{r^q} \right )
\nonumber \\
&&
- \left (\frac{v^{2s} {\dot r}^p}{r^{q+2}} \right ){\bf n} \times \left ( \left [ {\bf v} - \frac{3}{2}
{\dot r} {\bf n} \right ] \times (4 {\bf {\cal S}} + 3 \fourvec{\xi}) 
\right ) \,.
\label{timederiv1}
\eea
Another set of identities, to be used only in 3.5PN terms, require only the 
Newtonian equations of motion:
\bea
\frac{d}{dt} \left ( \frac{v^{2s} {\dot r}^p}{r^q} x^ix^j \right ) &=&
\frac{v^{2s-2} {\dot r}^{p-1}}{r^{q+1}}
\left \{  \left [pv^4 - (p+q)v^2{\dot r}^2 - 2s{\dot r}^2\frac{m}{r}
-pv^2\frac{m}{r} \right ] x^ix^j
\right .
\nonumber \\
&&
\left .
+2 v^2{\dot r} r x^{(i}v^{j)} \right \} \,,
\nonumber \\
\frac{d}{dt} \left ( \frac{v^{2s} {\dot r}^p}{r^q} v^iv^j \right ) &=&
\frac{v^{2s-2} {\dot r}^{p-1}}{r^{q+1}}
\left \{  \left [pv^4 - (p+q)v^2{\dot r}^2 - 2s{\dot r}^2\frac{m}{r}
-pv^2\frac{m}{r} \right ] v^iv^j
\right .
\nonumber \\
&&
\left .
-2m \frac{v^2{\dot r}}{r^2} x^{(i}v^{j)} \right \} \,,
\nonumber \\
\frac{d}{dt} \left ( \frac{v^{2s} {\dot r}^p}{r^q} x^iv^j \right ) &=&
\frac{v^{2s-2} {\dot r}^{p-1}}{r^{q+1}}
\left \{  \left [pv^4 - (p+q)v^2{\dot r}^2 - 2s{\dot r}^2\frac{m}{r}
-pv^2\frac{m}{r} \right ] x^iv^j
\right .
\nonumber \\
&&
\left .
+ v^2{\dot r} r \left ( v^iv^j - \frac{m}{r} n^in^j \right ) \right \} \,.
\label{timederiv2}
\eea

\section{Evolution of spins at 3.5PN order}
\label{app:spinevolve}

In this appendix we justify our assumption that
the individual spins are unaffected by
3.5PN spin-orbit effects, {\em i.e.} that 
$(d{\bf S}_1/dt)_{\rm 3.5PN-SO} =0$, and similarly for body 2.  First
we write down the general form that 3.5PN spin-orbit terms could take,
consistent with the assumptions used in earlier sections, namely
\bea
({\dot {\bf {\cal S}}}_1)_{\rm 3.5PN-SO} &=& \frac{\mu^2}{r^3}
\biggl \{ {\cal N}_1 {\bf {\cal S}}_1 
+ {\cal N}_2 {\bf n}{\bf n}\cdot {\bf {\cal S}}_1 
+ {\cal N}_3 {\dot r} {\bf n}{\bf v}\cdot {\bf {\cal S}}_1 
+ {\cal N}_4 {\dot r} {\bf v}{\bf n}\cdot {\bf {\cal S}}_1 
+ {\cal N}_5 {\bf v}{\bf v}\cdot {\bf {\cal S}}_1  
\biggr \} \,,
\label{s1dot}
\eea
where ${\cal N}_1$ and ${\cal N}_2$ are each linear combinations  
of $v^4$, $v^2 {\dot
r}^2$ and $v^2 m/r$ etc. 
at $O(\epsilon^2)$ (containing 6 terms each), and
${\cal N}_3$, ${\cal N}_4$ and ${\cal N}_5$ are each linear
combinations of $v^2$, ${\dot r}^2$ and $m/r$.  Note that other
possible terms, such as ${\bf L}_{\rm N} \times {\bf {\cal S}}_1$, or 
${\bf L}_{\rm N} {\bf L}_{\rm N} \cdot {\bf {\cal S}}_1$ can be rewritten as
linear combinations of the terms above.  

We now impose the physically reasonable
constraint ${\bf {\cal S}}_1 \cdot ({\dot {\bf {\cal
S}}}_1)_{\rm 3.5PN-SO} =0$, which implies that radiation reaction does
not change the magnitude of the body's spin, only its orientation.
That constraint implies that only the third and fourth terms in Eq.
(\ref{s1dot}) survive, and then only in an antisymmetric combination that
leaves $({\dot {\bf {\cal S}}}_1)_{\rm 3.5PN-SO}$
in the general form
\be
({\dot {\bf {\cal S}}}_1)_{\rm 3.5PN-SO} = 
\frac{\mu^2}{r^4} {\dot r}{\tilde {\bf L}}_N \times {\bf {\cal S}}_1
\left ( c_1 v^2 + c_2 {\dot r}^2 + c_3 \frac{m}{r} \right ) \,.
\label{s1dot2}
\ee
However, it is straightforward to show, using the identities in Appendix
\ref{app:timederiv}, that the right-hand side of Eq. (\ref{s1dot2}) 
can be written as a total time derivative and therefore can be
absorbed into ${\bf {\cal S}}_1$, independently of the values of
$c_1$, $c_2$ and $c_3$.  This is in accord with the result derived
from first principles in \cite{dire3}.

\begin{acknowledgements}
Supported in part by the National Science Foundation, Grant Nos.
PHY 03-53180 and PHY 06-52448, and by 
the National Aeronautics and Space Administration, Grant No.
NNG06GI60G.  CW is grateful to the Group
Gravitation Relativiste et Cosmologie (GR$\varepsilon$CO) of the Institut
d'Astrophysique de Paris for its hospitality while this work was being
completed.  We are also grateful to Bala Iyer for useful discussions during
an earlier (circa 1995) attack on this problem.
\end{acknowledgements}


\begin{thebibliography}{}
%
\bibitem{3min} Cutler, C., Apostolatos, T. A., Bildsten, L., Finn, L. S.,
Flanagan, \'E. E., Kennefick, D., Markovi\'c, D. M., Ori, A., Poisson, E., 
Sussman, G. J., Thorne, K. S.: Phys. Rev. Lett. {\bf 70}, 2984
(1993)

\bibitem{kww}
Kidder, L. E., Will, C. M., Wiseman, A. G.:
Phys. Rev. D {\bf 47}, R4183 (1993)

\bibitem{kidder}
Kidder, L. E.: 
Phys. Rev. D {\bf 52}, 821 (1995)

\bibitem{papapetrou1}
Papapetrou, A.: Proc. Roy. Soc. (London) {\bf 209A}, 248 (1951)

\bibitem{papapetrou2}
Corinaldesi, E., Papapetrou, A.: Proc. Roy. Soc. (London) {\bf 209A}, 259
(1951)

\bibitem{obrien} 
O'brien, G.: 
Gen. Rel. Grav. {\bf 10}, 129 (1979)

\bibitem{barkerocon1} 
Barker, B. M., O'Connell, R. F.: 
Phys. Rev. D {\bf 2}, 1428 (1970)

\bibitem{barkerocon2}
Barker, B. M., O'Connell, R. F.: 
Phys. Rev. D {\bf 12}, 329 (1975)

\bibitem{barkeroconrev}
Barker, B. M., O'Connell, R. F.: 
Gen. Rel. Grav. {\bf 11}, 149 (1979)

\bibitem{dire3} Will, C. M.:
Phys. Rev. D {\bf 71}, 084027 (2005)

\bibitem{dire4} Wang, H., Will, C. M.:
Phys. Rev. D {\bf 75}, 064017 (2007)

\bibitem{iyerwill1}
Iyer, B. R., Will, C. M.: 
Phys. Rev. Lett. {\bf 70}, 113 (1993)

\bibitem{iyerwill2}
Iyer, B. R., Will, C. M.: 
Phys. Rev. D {\bf 52}, 6882 (1995)

\bibitem{MTW}
See, for example, Sec. 36.8 of Misner, C. W., Thorne, K. S., Wheeler, J. A.:
Gravitation.
Freeman, San Francisco (1973)

\bibitem{DD81} Damour, T., Deruelle, N.: 
Phys. Lett. {\bf 87A}, 81 (1981)

\bibitem{damour300} Damour, T.: In 
Hawking, S. W., Israel, W. (eds.)  
300 Years of Gravitation.
Cambridge University Press,
Cambridge (1987), p. 128

\bibitem{gopuiyer2}
Gopakumar, A., Iyer, B. R., Iyer, S.:
Phys. Rev. D {\bf 55}, 6030 (1997); Erratum, {\it ibid} {\bf 57}, 6562
(1998).

\bibitem{barkeroconssc}
Barker, B. M., O'Connell, R. F.: 
Gen. Rel. Grav. {\bf 5}, 539 (1974)
\end{thebibliography}


\end{document}